\documentclass[elsarticle,onecolumn,nofootinbib,floatfix,12pt]{revtex4}
\usepackage{amssymb,amsfonts}
\usepackage{amsmath}
\usepackage{bm}

\newcommand{\xx}{\mathbf x}
\newcommand{\vv}{\mathbf v}
\newcommand{\pp}{\mathbf p}
\newcommand{\kk}{\mathbf k}
\newcommand{\BB}{\mathbf B}
\newcommand{\bb}{\mathbf b}
\newcommand{\om}{\bm \omega}

\newcommand{\fp}[2]{\frac{\partial #1}{\partial #2}}
\newcommand{\dfp}[2]{\dfrac{\partial #1}{\partial #2}}
\newcommand{\CF}{{\cal F}}
\newcommand{\CE}{{\cal E}}
\newcommand{\EE}{\bm \CE}
\newcommand{\avr}[1]{\langle #1\rangle_R}
\newcommand{\av}[1]{\langle #1\rangle}
\newcommand{\avl}[1]{\langle #1\rangle_L}
\newcommand{\be}{\begin{equation}}
\newcommand{\ee}{\end{equation}}
\newcommand{\bea}{\begin{eqnarray}}
\newcommand{\eea}{\end{eqnarray}}

\newcommand{\intp}{\frac{1}{\chi}\int_{\pp}}
\DeclareMathOperator\arcotanh{arcotanh}

\begin{document}

\title{Chiral heat wave in cold Fermi liquid and modified zero sound}

\author{D. Frenklakh$^{1,3}$,  A. Gorsky$^{2,3}$}

\address{$^1$Institute of Theoretical and Experimental Physics,\\$^2$Institute of Information Transmission Problems of the Russian Academy of Sciences,
Moscow, Russia, \\ $^3$Moscow Institute of Physics and Technology, Dolgoprudny 141700, Russia 
.}

\begin{abstract}
We discuss kinetic equations involving the anomalous terms responsible for the chiral 
anomaly. The general chiral heat wave in cold 
Fermi liquid is described and the modification of the anomalous zero sound at  small temperature
and vorticity is found.

\end{abstract}
\maketitle

\section{Introduction}

The anomalous contributions in the kinetic equations and hydrodynamics
have been overlooked for many years however now the corresponding
modifications are well established \cite{hydro1,kin1} in both settings.
The anomalous terms follow from the nontrivial connection in the 
momentum space emerging due to the level crossing phenomena. For the
relativistic particle the Berry connection involves  the
monopole connection where helicity yields the corresponding charge 
of the particle. Both chiral anomaly in the external magnetic field 
and the mixed anomaly modify the kinetic equations \cite{kin1,mixed}.
Recently the second order approximation to the kinetic and 
hydrodynamical equations in the chiral matter has been considered \cite{gorbar}.

The new terms in the kinetics imply the possibility of the new types 
of the collective excitations and modification of the old ones. The first 
example of the new modes was developed in \cite{khar} for the chiral magnetic
wave and later the similar chiral vortical wave has been found \cite{vort}.
The mixing of these new modes as well as the modes emerging from the
temperature gradients have been discussed in \cite{stephanov,chernodub,murchikova}. The most 
general  pattern of the mixing considered so far has been analyzed in \cite{frenklakh}.

The analysis takes some care due to the subtleties concerning the 
Lorentz invariance \cite{lor1,lor2,lor3}. In particular it was argued
that condition of the Lorentz invariance imposes nontrivial restriction 
on the dynamics of the collective modes and for instance the careful
analysis shows the way how the total contribution for the chiral magnetic effect
comes from the several independent contributions \cite{lor3}.

In this Letter we focus on the collective wave in the cold Fermi liquid 
\cite{yamamoto} assuming
that temperature is small enough. We investigate the mixing of the different 
modes in terms of the kinetic equations in two opposite limits in the relaxation 
time. In the first limit we consider the hydrodynamical approximation and reproduce
the dispersion law of the collective modes derived in \cite{frenklakh} in the
different approach. In the opposite limit we consider the zero sound corresponding
to the fluctuations of the Fermi surface. The modification of the anomalous
zero sound discussed in \cite{gor, stephanov1} due to the vorticity and temperature
will be derived. 

The note is organized as follows. First, we remind the kinetic equations involving the
anomalous terms. In Section 3 we consider the hydrodynamic limit of the kinetic 
equations while in Section 4 we derive the modifications of the anomalous zero sound
in the cold Fermi liquid.

\section{Kinetic description}

A natural way to describe  Fermi liquid is through right and left quasiparticles. We denote their  distribution functions as $n_R(t,\xx,\pp)$ and $n_L(t,\xx,\pp)$, respectively. Energy of  the quasiparticles is given by the energy functionals $\epsilon_R(\pp)[n_R,n_L]$ and $\epsilon_L(\pp)[n_R,n_L]$. Later we always refer to the quasiparticles as particles, because we will not deal with any actual particles. We begin with usual kinetic equations for both kinds of particles:

\be\label{Kin_eq}
\fp{n_{R/L}}{t}+\dot{\xx}_{R/L}\cdot\fp{n_{R/L}}{\xx}+\dot{\pp}_{R/L}\cdot\fp{n_{R/L}}{\pp} = C_{R/L}[n_R,n_L],
\ee

where $C_{R/L}$ are collision integrals. They are supplemented with the equations of motion which take into account rotation of the liquid (treated as a perturbation to the metric) and anomalous Berry curvature contribution. The gravitomagnetic analogy tells us that for small angular velocities quasiparticles in rotating background behave as if they were in an external energy-dependent magnetic field $\BB\sim 2\epsilon\om$, where $\epsilon$ is a quasiparticle energy (\cite{basar}). The equations of motion for quasiparticles with Berry curvature in an external magnetic field are known (\cite{yamamoto}), so in our case they are: 

\bea\label{motion}
\sqrt{G_R}\dot{\xx}_R = \vv_R + 2\epsilon_R\om(\vv_R\cdot\bb_R) + \EE_R\times\bb_R, \\
\sqrt{G_L}\dot{\xx}_L = \vv_L + 2\epsilon_L\om(\vv_L\cdot\bb_L) + \EE_L\times\bb_L, \\
\sqrt{G_R}\dot{\pp}_R = \EE_R +2\epsilon_R\vv_R\times\om + (\EE_R\cdot\om)2\epsilon_R\bb_R, \\
\sqrt{G_L}\dot{\pp}_L = \EE_L +2\epsilon_L\vv_L\times\om + (\EE_L\cdot\om)2\epsilon_L\bb_L,
\eea

where $\vv_{R/L} = \dfp{\epsilon_{R/L}}{\pp}$ are particle velocities, $p = |\pp|$, $v_{F}$ is Fermi speed(we assume that it is the same for left and right particles), $\EE_{R/L} = - \dfp{\epsilon_{R/L}}{\xx}$, $\bb_{R/L} = \pm\dfrac{\hat{\pp}}{2p^3}$ are Berry connections in momentum space and factors $\sqrt{G_{R/L}} = 1 + 2\epsilon_{R/L}(\bb_{R/L}\cdot\om)$ modify phase space volume due to the interplay between Berry phase and vorticity.

We are interested in waves propagating in the liquid so we parametrize the distribution functions as planar wave fluctuations above the equilibrium configuration (which we assume to be the same for left and right particles):

\be
n_{R/L} = n^0 + \fp{n^0}{\mu}h_{R/L}(\pp)e^{i(\nu t - \kk\xx)},
\ee

where $n^0 = \dfrac{1}{e^{\beta(\epsilon^0 -\mu)} + 1}$ is usual Fermi-Dirac distribution (note that $\epsilon^0$ are energies of particles having this particular distribution function). Throughout this paper we consider $\mu\gg T$ which corresponds to cold regime.

Such parametrization of the distribution functions leads to parametrization of energy fluctuations and 
collision integral as

\be
\delta\epsilon_{R/L} = \epsilon_{R/L} - \epsilon^0 = \CF_{R/L}[h_R,h_L] e^{i(\nu t - \kk\xx)} + O(h^2),
\ee

and, since in equilibrium the collision integral is zero,

\be\label{coll_lin}
C_{R/L} = \fp{n^0}{\mu}I_{R/L}[h_R,h_L]e^{i(\nu t - \kk\xx)} + O(h^2),
\ee

where $\CF$'s and $I$'s are linear functionals. Plugging all these (\ref{motion})-(\ref{coll_lin}) back into the kinetic equations (\ref{Kin_eq}) after some rearrangement we obtain:

\bea\label{lin_kin_R}
-i\nu h_R + \dot{\xx}^0_R\cdot\left(i\kk + 2\frac{\epsilon^0}{\sqrt{G_R}}\om\times\fp{~}{\pp}\right)\left\{h_R+\CF_{RR}[h_R]+\CF_{RL}[h_L]\right\} = I_R[h_R,h_L], \\
-i\nu h_L + \dot{\xx}^0_L\cdot\left(i\kk + 2\frac{\epsilon^0}{\sqrt{G_L}}\om\times\fp{~}{\pp}\right)\left\{h_L+\CF_{LR}[h_R]+\CF_{LL}[h_L]\right\} = I_L[h_R,h_L], \label{lin_kin_L}
\eea

where we have split the linearized energy functionals into left and right parts: $\CF_R[h_R,h_L] = \CF_{RR}[h_R]+\CF_{RL}[h_L]$ and  $\CF_L[h_R,h_L] = \CF_{LR}[h_R]+\CF_{LL}[h_L]$. Here $\dot{\xx}^0_{R/L} = \dfrac{1}{\sqrt{G_{R/L}}}[\vv^0 + 2\epsilon_{R/L}\om(\vv^0\cdot\bb_{R/L})]$.

Further we are going to study two opposite regimes : hydrodynamic regime of low frequency and small wave vector and zero sound regime of high frequency and large wave vector. If $\tau$ is some characteristic relaxation time entering the collision integral, these regimes are given, respectively, by $\nu\tau \ll 1$ and $\nu\tau\gg 1$.

\section{Hydrodynamic regime}

In hydrodynamic regime $\nu,\kk\rightarrow 0 $. Note that there exist solutions to the kinetic equations given by infinitesimal constant shift of chemical potentials and temperature, accompanied by the corresponding shifts in the functions of particle energy in equilibrium (the last is because for Fermi liquid energy of particles is, generally speaking, a functional of the distribution function). So we parametrize the fluctuations as (note that we allow the chemical potentials of right and left particles to fluctuate independently):

\be
h_{R/L} = \delta\mu_{R/L} - \delta\epsilon_{R/L} - \frac{\epsilon^0 - \mu}{T}\delta T
\ee

Since this is a solution for small constant $\delta\mu_{R/L}$ and $\delta T$, we expect that to the lowest order in $\nu$ and $\kk$ the solution we are looking for will have the same form. From here on let us assume the simplest possible form of the linearized energy functionals: $\CF_{RR}[h_R] = F_S\avr{h_R}$,  $\CF_{RL}[h_L] = F_A\avl{h_L}$, $\CF_{LR}[h_R] = F_A\avr{h_R}$ and $\CF_{LL}[h_L] = F_S\avl{h_L}$, where we have introduced averaging over momentum space:

\bea
\avr{...} = \intp\sqrt{G_R}~\fp{n^0}{\mu}(...) ,\\
\avl{...} = \intp\sqrt{G_L}~\fp{n^0}{\mu}(...).
\eea

Here $\int_{\pp} = \int \dfrac{d^3p}{(2\pi)^3}$, $\chi = \int_{\pp}\sqrt{G_R}~\dfp{n^0}{\mu} = \int_{\pp}\sqrt{G_L}~\dfp{n^0}{\mu}$ since the equilibrium distribution is isotropic and the differences between $\sqrt{G_R}$ and $\sqrt{G_L}$ do not show up. Note that the normalization is so that $\avr{1} = \avl{1} = 1$. So energy fluctuations have the form

\bea
\delta\epsilon^0_R = F_S\avr{h_R} + F_A\avl{h_L},\\
\delta\epsilon^0_L = F_A\avr{h_R} + F_S\avl{h_L}.
\eea

Let us for convenience further denote $\delta\mu_R$ as $h_1$, $\delta\mu_L$ as $h_2$ and $(-\delta T)$ as $h_3$. 

To get rid of the collision integrals let us make use of conservation of the number of right and left particles and energy. It implies for collision integrals

\bea
\int_{\pp}\sqrt{G_R}C_R[n_R,n_L] = 0 ,~~~~~~~~~~~~~~~~~~~~~ \\
\int_{\pp}\sqrt{G_L}C_L[n_R,n_L] = 0 ,~~~~~~~~~~~~~~~~~~~~~ \\
\int_{\pp}(\sqrt{G_R}\epsilon_R[n_R,n_L] C_R[n_R,n_L] + \sqrt{G_L}\epsilon_L[n_R,n_L] C_L[n_R,n_L]) = 0,
\eea

for any $n_R$, $n_L$ and corresponding $\epsilon_R[n_R,n_L]$, $\epsilon_L[n_R,n_L]$ which implies for any $h_R$, $h_L$:

\bea
\avr{I_R[h_R,h_L]} = 0 ,~~~~~~~~~~~~ \\
\avl{I_L[h_R,h_L]} = 0 ,~~~~~~~~~~~~ \\
\avr{\epsilon_R I_R[h_R,h_L]} + \avl{\epsilon_L I_L[h_R,h_L]} = 0.
\eea

So we act on the equations (\ref{lin_kin_R}),(\ref{lin_kin_L}) with the averaging operation to obtain

\bea\label{sys}
-i\nu\av{h_R} + i\kk\cdot\avr{\dot{\xx}^0_R (h_R + F_S\av{h_R} + F_A\av{h_L})} = 0,~~~~~~~~~~~~~~~~~~~~~~~~~~~~~ \\
-i\nu\av{h_L} + i\kk\cdot\avl{\dot{\xx}^0_L (h_L + F_A\av{h_R} + F_S\av{h_L})} = 0,~~~~~~~~~~~~~~~~~~~~~~~~~~~~~ \\
-i\nu(\av{h_R\epsilon^0} + \av{h_L\epsilon^0}) + i\kk\cdot(\avr{\dot{\xx}^0_R (h_R+ F_S\av{h_R} + F_A\av{h_L})\epsilon^0} + \avl{\dot{\xx}^0_L (h_L + F_A\av{h_R} + F_S\av{h_L})\epsilon^0}) = 0.~~~ \label{sys_last}
\eea

Here we used the fact that since the quantities such as $h_R$, $h_L$, $h_R\epsilon^0$, $h_L\epsilon^0$ are isotropic we may forget about the $\sqrt{G}$'s factors and average them just as

\be
\av{...} = \intp \fp{n^0}{\mu}.
\ee

Also this isotropy leads to vanishing of the terms, proportional to $\om\times\dfp{~}{\pp}$. Note that since $h$'s are small and $\epsilon_{R/L} = \epsilon^0 + "terms~ linear~ in~ h"$ in the equation (\ref{sys_last}) we only keep $\epsilon^0$ since we are solely interested in the terms, linear in $h$. Let us for future convenience change the notation here. From here on we will work with the variables $h_V = h_1+ h_2$ and $h_A = h_1-h_2$ having the meaning of fluctuations of vector and axial chemical potentials correspondingly. Also for convenience in future we will not work with the two first equations of the system (\ref{sys}) separately but instead we will sum them up and subtract the second from the first one. 

Then,

\bea
\av{h_R} + \av{h_L}  = h_V - (F_S+F_A)(\av{h_R} + \av{h_L}) + 2\av{\frac{\epsilon^0 - \mu}{T}}h_3 , \\
\av{h_R} - \av{h_L}  = h_A + (F_A - F_S)(\av{h_R} - \av{h_L}) ,~~~~~~~~~~~
\eea

so

\bea
\av{h_R+h_L} = \frac{h_V + 2\av{\frac{\epsilon^0 -  \mu}{T}h_3}}{1+F_S+F_A} , \\
\av{h_R-h_L} = \frac{h_A}{1+F_S-F_A}.~~
\eea

Similarly 

\be
\av{(h_R + h_L)\epsilon^0} = [h_V - (F_S+F_A)\av{h_R + h_L}]  \av{\epsilon^0} +  2\av{\frac{\epsilon^0 - \mu}{T}\epsilon^0}h_3 .
\ee

Now we calculate $\av{\epsilon^0}$ and $\av{(\epsilon^0)^2}$. Since the temperature is low, $T\ll\mu$, we will only keep terms up to quadratic in temperature. At low temperatures all the excitations are localized near the Fermi sphere, so we will expand the dispersion relation of the particles near it :

\be
\epsilon^0 = \mu + v_F(p-p_F) ,
\ee

Here $p_F$ is Fermi momentum and we assume $p_F v_F\sim\mu$. In fact, the next terms with the second and the third powers of $p-p_F$ would also contribute to some quantities we are going to calculate in the order, interesting to us. However, it doesn't seem that including such terms will give rise to any conceptual difference. So we will omit them to reduce the mess in the calculations and restrict ourselves with the linear dispersion relation for the particles not far from the Fermi surface. Reversing this relation we obtain

\be
p = p_F + \frac{\epsilon^0 - \mu}{v_F}.
\ee

Then,

\bea
\nonumber  \av{\epsilon^0} = \intp\fp{n^0}{\mu}\epsilon^0 = \frac{1}{2\pi^2\chi}\int_0^{\infty} \frac{\beta\epsilon^0 e^{\beta(\epsilon^0 - \mu)}}{(e^{\beta(\epsilon^0 - \mu)} + 1)^2}p^2 dp \approx \frac{T}{2\pi^2\chi v_F}\int_{-\infty}^{\infty}\frac{(x+\beta\mu)e^x}{(e^x+1)^2}\left(p_F^2 + \frac{2p_F Tx}{v_F} + \frac{x^2 T^2}{v_F^2} \right)dx \\ 
 \approx \frac{1}{2\pi^2\chi v_F}\left(\mu p_F^2 + \frac{T^2\pi^2}{3v_F^2}(2p_Fv_F + \mu)\right) .~~~~~~~~~~~~~~~~~~~~~~~~~~~~~~~~~~~~~~
\eea

 Note that when integrating over $x = \beta(\epsilon^0 - \mu)$ we have set the lower integration limit to minus infinity, since it is actually $-\beta\mu$ where the function is exponentially small and does not contribute to the integral. Similarly,

\be
\av{(\epsilon^0)^2} \approx \frac{1}{2\pi^2\chi v_F}\left[ p_F^2\mu^2 + \frac{T^2\pi^2}{3}\left(p_F^2 + 4\frac{\mu}{v_F}p_F + \frac{\mu^2}{v_F^2} \right) + \frac{7\pi^4 T^4}{15~v_F^2} \right] .
\ee

Let us also present the expression for $\chi$ :

\be
\chi = \int_{\pp} \fp{n^0}{\mu} = \frac{1}{2\pi^2 v_F}\left(p_F^2 + \frac{T^2\pi^2}{3v_F^2}
\right).
\ee

Then,

\bea
\av{\frac{\epsilon^0 - \mu}{T}} = \frac{p_F T}{3\chi v_F^2}~~,~~~~~~~~~~ \\
\av{\frac{\epsilon^0-\mu}{T}\epsilon^0}= \frac{T}{6\chi v_F}\left(p_F^2  + 2p_F\frac{\mu}{v_F} + \frac{7\pi^2 T^2}{5~v_F^2}\right).
\eea

So, finally

\bea
\av{h_R+h_L} = \frac{h_V + h_3\frac{2p_F T}{3\chi v_F^2}}{1+F_S+F_A}, ~~~~~~~~~~~~~~~~~~~~~~~~~~~~~~~~~~~~~~~~~~~~~~~~~~~\\
\av{h_R-h_L} = \frac{h_A}{1+F_S-F_A} ,~~~~~~~~~~~~~~~~~~~~~~~~~~~~~~~~~~~~~~~~~~~~~~~~~~~ \\
\av{(h_R+h_L)\epsilon^0} = \frac{h_V}{1+F_S+F_A}\left(\mu+\frac{2T^2\pi^2}{3p_F v_F}\right) + \frac{h_3 T}{3\chi v_F}\left(p_F^2 + \frac{2p_F\mu}{v_F(1+F_S+F_A)} + \frac{\pi^2 T^2(21+F_S+F_A)}{15~v_F^2(1+F_S+F_A)}\right).
\eea

Now let's proceed to calculating terms like $\avr{\dot{\xx}^0_R h_R}$ and $\avr{\dot{\xx}^0_R\epsilon^0 h_R}$. Since the only term in $h$'s depending on $p$ is $\dfrac{\epsilon^0-\mu}{T}$, it will suffice to calculate $\avr{\dot{\xx}^0_R}$ and $\avr{\dot{\xx}^0_R(\epsilon^0-\mu)}$. Let us remind that $\dot{\xx}^0_{R/L} = \dfrac{1}{\sqrt{G_{R/L}}}[\vv^0 + 2\epsilon_{R/L}\om(\vv^0\cdot\bb_{R/L})]$. Note that the first term will not give a contribution after averaging since the rest of the expression is isotropic.

\be
\avr{\dot{\xx}_R^0} = \frac{2\om}{2\pi^2\chi}\int_0^{\infty}\epsilon^0 v^0 \frac{1}{2p^2}p^2 dp \fp{n^0}{\mu} = \frac{\om}{2\pi^2\chi}\int^{\infty}_{-\infty}\frac{T(x+\beta\mu)v_F \beta e^x}{(e^x+1)^2}\frac{T}{v_F}dx = \frac{\om \mu}{2\pi^2\chi}.
\ee

Similarly,

\be
\avr{\dot{\xx}_R^0(\epsilon^0-\mu)} = \frac{\om T^2}{6\chi} .
\ee

For the left particles all that changes compared to the right ones is $\bb_R\rightarrow \bb_L = -\bb_R$, which effectively just causes terms with $\om$ to change sign. So, we obtain

\bea
\avr{\dot{\xx}_R^0(h_R+F_S\av{h_R} + F_A\av{h_L})} + \avl{\dot{\xx}_L^0(h_L+F_A\av{h_R}+F_S\av{h_L})} = \frac{\mu\om}{2\pi^2\chi}h_A,~~~~~~~~ \\
\avr{\dot{\xx}_R^0(h_R+F_S\av{h_R} + F_A\av{h_L})} - \avl{\dot{\xx}_L^0(h_L+F_A\av{h_R}+F_S\av{h_L})} = \frac{\mu\om}{2\pi^2\chi}h_V + \frac{\om T}{3\chi}h_3, ~~~ \\
\avr{\dot{\xx}^0_R (h_R+F_S\av{h_R} + F_A\av{h_L})\epsilon^0} + \avl{\dot{\xx}_L^0 (h_L+F_A\av{h_R}+F_S\av{h_L})\epsilon^0} = \frac{\om}{\chi}\left(\frac{\mu^2}{2\pi^2} + \frac{T^2}{6}\right)h_A.
\eea

So, the system (\ref{sys}) - (\ref{sys_last}) transforms into

\bea
\nu\left(\frac{h_V}{F_1} + h_3 \frac{2p_F T}{3\chi v_F^2 F_1}\right) - \frac{(\kk\cdot\om)}{\chi} h_A \frac{\mu}{2\pi^2} = 0 ,~~~~~~~~~~~~~~~~~~~~~~~~~~~~~~~~ \\
\nu \frac{h_A}{F_2} -\frac{(\kk\cdot\om)}{\chi} \left(h_V \frac{\mu}{2\pi^2} + \frac{h_3 T}{3}\right) = 0,~~~~~~~~~~~~~~~~~~~~~~~~~~~~~~~~~~ \\
\nu\left[\frac{h_V \mu}{F_1}\left(1 + \frac{2\pi^2 T^2}{3 \mu p_F v_F}\right) + \frac{h_3 T}{3\chi v_F}\left(p_F^2 + \frac{2p_F\mu}{v_F F_1} + \frac{\pi^2 T^2(20 + F_1)}{15~v_F^2 F_1}\right)\right] - \frac{h_A(\kk\cdot\om)}{\chi}\left(\frac{\mu^2}{2\pi^2 } + \frac{T^2}{6}\right) = 0.~~~
\eea

Here we have introduced $F_1 = 1+F_S+F_A$, $F_2 = 1+F_S-F_A$. For this system to be consistent the following dispersion relation should be true (so that it's determinant is zero): 

\be \label{ans_CHVW}
\nu =\pm\frac{\mu(\kk\cdot\om)}{2\pi^2\chi}\sqrt{F_1 F_2} \left[1 + \frac{T^2\pi^2}{6\mu^2 F_1}\left(1 - 4A + 4A^2\right)\right]
\ee

Here we have introduced dimensionless parameter $A = \dfrac{\mu}{v_Fp_F}$.

At the first sight, at zero temperature this dispersion relation differs from the known expression for the Chiral Vortical Wave (first obtained in \cite{vort}):

\be \label{ans_CVW}
\nu = \pm\frac{\mu(\kk\cdot\om)}{2\pi^2\chi}.
\ee
 
However, we should take into account that, firstly, our definition of $\chi$ differs from the one in \cite{vort}, and secondly, that in \cite{vort} it was supposed, that vector and axial susceptibilities are the same, which is not true in our case(if $F_A\neq 0$). Namely, our definition of $\chi$(which we will later refer to $\chi_{our}$) is given by $\chi_{our} = \int_{\pp}\sqrt{G_R}~\dfp{n^0}{\mu} = \int_{\pp}\dfp{n^0}{\mu}$ which is not exactly susceptibility $\chi = \dfrac{\delta N^0}{\delta\mu}$ as in \cite{vort} (where $N^0$ is density of particles) due to the interactions. As to the second point, if one does not suppose $\chi_V = \chi_A$, where $\chi_V = \dfrac{\delta N_V}{\delta\mu_V}$ and $\chi_A = \dfrac{\delta N_A}{\delta\mu_A}$ (here $N_{A/V}$ are vector and axial charge densities respectively and $\mu_{V/A}$ are vector and axial chemical potentials respectively), then in the expression (\ref{ans_CVW}) $\chi$ (denoting both $\chi_V$ and $\chi_A$ in case they coincide) should be replaced with $\sqrt{\chi_V \chi_A}$. In order to show that (\ref{ans_CHVW}) at zero temperature indeed coincides with (\ref{ans_CVW}) we will derive the expressions for $\chi_V$ and $\chi_A$ via $\chi_{our}$, $F_1$ and $F_2$:

\be
\chi_V = \frac{\delta N_V}{\delta\mu_V} = \frac{\delta N_R + \delta N_L}{\delta\mu_V},
\ee

Since

\be
\delta N_R = \int_{\pp}\fp{n_R}{\mu_R}\delta h_R, ~~~~~~~~~~ \delta N_L = \int_{\pp}\fp{n_L}{\mu_L}\delta h_L,
\ee

where

\be
\delta h_R = \delta\mu_R - \delta\epsilon_R, ~~~~~~~~ \delta h_L = \delta\mu_L - \delta\epsilon_L,
\ee
 
and

\be 
\delta\epsilon_R = F_S\delta h_R + F_A\delta h_L, ~~~~~~~~~ \delta\epsilon_L = F_A\delta h_R + F_S\delta h_L,
\ee

we get

\be
\delta N_R = \int_{\pp}\fp{n^0}{\mu}\frac{\delta\mu_R}{F_1}, ~~~~~~~~~~ \delta N_L = \int_{\pp}\fp{n^0}{\mu}\frac{\delta\mu_L}{F_1}, 
\ee
 
and finally

\be
\delta N_V = \int_{\pp}\fp{n^0}{\mu}\frac{\delta\mu_R + \delta\mu_L}{F_1} = \int_{\pp}\fp{n^0}{\mu}\frac{\delta\mu_V}{F_1},
\ee

which gives

\be 
\chi_V = \frac{\chi_{our}}{F_1}.
\ee 
 
Similar calculation gives

\be 
\chi_A = \frac{\chi_{our}}{F_2},
\ee
 
which confirms that our answer (\ref{ans_CHVW}) coincides with (\ref{ans_CVW}) at zero temperature. 
 
At non-zero temperatures the corrections quadratic in temperature represent the mixing of Chiral Heat and Vortical Waves (\cite{chernodub, frenklakh}).

\section{Zero sound}

\subsection{Zero temperature}

The regime opposite to the hydrodynamic one is given by $\nu\tau\gg 1$, so we may neglect collision terms. Then kinetic equations (\ref{lin_kin_R}) and (\ref{lin_kin_L}) look like

\bea \label{zeros1}
-i\nu h_R + \dot{\xx}^0_R\cdot\left(i\kk + 2\epsilon^0\om\times\fp{~}{\pp}\right)\left\{h_R+\CF_{RR}[h_R]+\CF_{RL}[h_L]\right\} = 0, \\ \label{zeros2}
-i\nu h_L + \dot{\xx}^0_L\cdot\left(i\kk + 2\epsilon^0\om\times\fp{~}{\pp}\right)\left\{h_L+\CF_{LR}[h_R]+\CF_{LL}[h_L]\right\} = 0,
\eea

At first let us work at zero temperature and later we will find the temperature corrections perturbatively. For simplicity we are going to analyze excitations propagating along the vorticity $\kk ||\om$. Let us look for the axially-symmetric solutions, so that $h_R = h_R(\theta)$, $h_L = h_L(\theta)$, where $\theta$ is the angle between $\pp$ and $\kk$. Under these assumptions the terms of the form $\om\times\dfp{~}{\pp}$ will vanish in the above equations. We will also assume, like in the previous subsection, that the linearized energy functionals have the simplest possible form $\CF_{RR}[h_R] = F_S\avr{h_R}$,  $\CF_{RL}[h_L] = F_A\avl{h_L}$, $\CF_{LR}[h_R] = F_A\avr{h_R}$ and $\CF_{LL}[h_L] = F_S\avl{h_L}$. Performing the averaging in the axially-symmetric case we obtain

\bea
\CF_{RR}[h_R] = \frac{F_S}{\chi}\int_{\pp}\fp{n^0}{\mu}\sqrt{G_R}h_R(\theta)  = \frac{F_S}{4\pi^2\chi}\int_0^{\pi}\sin\theta h_R(\theta) d\theta\int_0^{\infty}p^2dp\frac{\beta e^{x}}{e^x+1}\left(1+\frac{2\epsilon^0\omega\cos\theta}{2p^2}\right) \nonumber \\
 = F_S\left(\int_0^{\pi}\frac{\sin\theta h_R(\theta)d\theta}{2} + \frac{\mu\omega}{4\pi^2\chi v_F}\int_0^{\pi}\sin\theta\cos\theta h_R(\theta) d\theta\right).
\eea

Let us denote 

\be
B_{R/L}^0 = \int_0^{\pi}\frac{\sin\theta h_{R/L}(\theta) d\theta}{2}~,~~~~~~~~~~~~~~~~~~ D_{R/L}^0 = \int_0^{\pi}\frac{\sin\theta \cos\theta h_{R/L}(\theta) d\theta}{2}, 
\ee
\be
a_0 = \frac{\mu\omega}{2\pi^2\chi v_F}|_{T=0} = \frac{\mu\omega}{p_F^2}
\ee

Then

\bea
\CF_{RR}[h_R(\theta)] = F_S(B_R^0 + a_0 D_R^0) ~,~~~~~~~~ \CF_{RL}[h_L(\theta)] = F_A(B_L^0 - a_0 D_L^0) ,\\
\CF_{RL}[h_R(\theta)] = F_A(B_R^0 + a_0 D_R^0) ~,~~~~~~~~ \CF_{LL}[h_L(\theta)] = F_S(B_L^0 - a_0 D_L^0) .
\eea

Taking all this into account and plugging back into (\ref{zeros1})-(\ref{zeros2}) we obtain

\bea
-s~ h_R(\theta)\left(1+a_0\cos\theta\right) + \left(\cos\theta+a_0\right)[h_R(\theta)+F_S(B_R^0+a_0 D_R^0) + F_A(B_L^0-a_0 D_L^0)] = 0 ,\\
-s~ h_L(\theta)\left(1-a_0\cos\theta\right) + \left(\cos\theta-a_0\right)[h_R(\theta)+F_A(B_R^0+a_0 D_R^0) + F_S(B_L^0-a_0 D_L^0)] = 0.
\eea

Here we denoted $s = \dfrac{\nu}{v_F k}$. Expressing $h_{R/L}^0$ in terms of $B_{R/L}^0$, $D_{R/L}^0$ and plugging it back into the definitions of $B_{R/L}^0$, $D_{R/L}^0$ we get the closed system:

\bea \label{sys_0_zeros1}
A_R^0 = I(a_0) (F_S A_R^0 +  F_A A_L^0) ,\\  \label{sys_0_zeros2}
A_L^0 = I(-a_0) (F_A A_R^0 + F_S A_L^0) ,
\eea

where we denoted $A_{R/L}^0 = B_{R/L}^0 \pm a_0 D_{R/L}^0 $ and introduced the following integral

\be
I(a_0) = \frac{1}{2}\int_{-1}^1\frac{dx(1+a_0 x)(x+a_0)}{s-x+a_0(sx-1)}.
\ee

The determinant of the system (\ref{sys_0_zeros1})-(\ref{sys_0_zeros2}) should be zero for it to be consistent, which gives us an equation upon $I(a_0)$, $I(-a_0)$ and, therefore, $s$:

\be \label{zeros_fin_0}
I(a_0)I(-a_0) (F_S^2 - F_A^2) - F_S[I(a_0) + I(-a_0)] + 1 = 0.
\ee

Note that we assume that the vorticity is small, which implies $a_0\ll 1$ and we may calculate $I(a_0)$ only up to the lowest orders in $a_0$. It follows from  (\ref{zeros_fin_0}) that terms linear in $a_0$ will not contribute, so we keep terms up to the second order.  Parametrizing the integral as

\be
I(a_0) = L_0(s_0) + a_0 L_1(s_0) + a_0^2 L_2(s_0), 
\ee

we easily find that

\bea
L_0(s_0) = s_0\arcotanh s_0 -1, ~~~~~~~~~~~ \\
L_1(s_0) = 3s_0(s_0\arcotanh s_0 -1),~~~~~~~~ \\
L_2(s_0) = 2s_0[-3s_0 + (3s_0^2 - 1)\arcotanh s_0].~
\eea

Here we changed the notation from $s$ to $s_0$ to emphasize that it is the solution at zero temperature. The equation (\ref{zeros_fin_0}) is now transformed into

\be
[L_0^2 - (L_1^2-2L_0L_2)a_0^2](F_S^2 - F_A^2) - 2F_S(L_0+a_0^2 L_2) + 1 = 0
\ee

By resolving this equation in the zero order in $a_0$ we find that $s_0$, which we will call in this order $s_0^0$ has to satisfy the irrational equation

\be  
\arcotanh s_0^0 = \frac{1}{s_0^0}\left(\frac{1}{F_S\pm F_A} + 1\right),
\ee

which is unmodified zero sound dispersion relation in case of two fermion species. For the quadratic in $a_0$ correction which actually contains the information about the vortical modification we have

\be
\delta s_0 = s_0- s_0^0 = -\mp a_0^2 \frac{F_S L_2(s_0^0) + [L_1(s_0^0)^2 - 2L_0(s_0^0)L_2(s_0^0)](F_S^2-F_A^2)]}{F_A [\arcotanh s_0^0 - \frac{s_0^0}{2((s_0^0)^2 - 1)}] }
\ee

These last two equations completely define the modified zero sound dispersion relation at zero temperature in the lowest order in vorticity (remember that $a_0$ is linear in $\omega$, so the corrections are quadratic in $\omega$).

\subsection{Non-zero temperature}

Now let us find thermal corrections to the modified zero sound dispersion relation. To do that, we will introduce a new term in the fluctuation of distribution function: it will be still axially-symmetric but there will be some dependence on the absolute value of momentum now : 

\be
h_{R/L} = h_{R/L}(\theta) + \delta h_{R/L} (p,\theta)
\ee

Here we assume that the second term is small compared to the first one (since it is related to the temperature it is of order of some power of $\dfrac{T}{\mu}$). Since both temperature and vorticity are small we will neglect terms of the form $\delta h_{R/L} \omega $. In this case 

\bea
\CF_{RR}[h_R] = F_S\avr{h_R(\theta)+\delta h_R} = F_S\left(A_R^0  - \frac{\omega\mu \pi^2 T^2}{3p_F^4 v_F^2}D_R^0 + \delta A_R\right) = F_S(A_R^0 + \delta a D_R^0 + \delta A_R) ,\\
\CF_{RL}[h_L] = F_A(A_L^0 - \delta a D_L^0 + \delta A_L) , ~~~~~~~~~~~~~~~~~~~~~~~~~~~~~~~~~\\
\CF_{LR}[h_R] = F_A(A_R^0 + \delta a D_R^0 + \delta A_R) , ~~~~~~~~~~~~~~~~~~~~~~~~~~~~~~~~~ \\
\CF_{LL}[h_L] = F_S(A_L^0 - \delta a D_L^0 + \delta A_L). ~~~~~~~~~~~~~~~~~~~~~~~~~~~~~~~~~
\eea
Here we denoted $- \dfrac{\omega\mu \pi^2 T^2}{3p_F^4 v_F^2}$ as $\delta a$, $\avr{\delta h_R}$ as  $\delta A_R$ and $\avl{\delta h_L}$ as $\delta A_L$.
When we plug these into (\ref{zeros1}) - (\ref{zeros2}) and note, that terms arising from $\om\times\dfp{~}{\pp}$ either vanish as before or are negligible in our approximation (we do not try to keep track of terms linear in $\delta h_{R/L}$ multiplied by some power of $\omega$), we obtain:

\bea
-s(h_R(\theta) + \delta h_R)\left(1+\frac{\epsilon_0\omega}{p^2}\cos\theta\right) + \left( \cos\theta + \frac{\epsilon_0\omega}{p^2}\right) \cdot \nonumber ~~~~~~~~~~\\
\cdot[h_R(\theta) + \delta h_R + F_S(A_R^0 + \delta a D_R^0 + \delta A_R) + F_A(A_L^0 - \delta a D_L^0 + \delta A_L)] = 0 , \\
-s(h_L(\theta) + \delta h_L)\left(1 - \frac{\epsilon_0\omega}{p^2}\cos\theta\right) + \left( \cos\theta - \frac{\epsilon_0\omega}{p^2}\right)\cdot \nonumber ~~~~~~~~~~\\
\cdot [h_L(\theta) + \delta h_L + F_A(A_R^0 + \delta a D_R^0 + \delta A_R) + F_S(A_L^0 - \delta a D_L^0 + \delta A_L)] = 0 .
\eea

Using these equations we express $\delta h_{R/L}$ via everything else (note that we know the expressions for $h_{R/L}(\theta)$) and then averaging the corresponding expressions  we find that the resulting system is consistent provided that $a_0$ in the discussion of zero-temperature case above is shifted to $a_0 + \delta a$ where $\delta a$ was defined above as $- \dfrac{\omega\mu \pi^2 T^2}{3p_F^4 v_F^2}$. Doing that exploits the fact that 

\be
\avr{\frac{\epsilon_0\omega}{p^2}} = \avl{\frac{\epsilon_0\omega}{p^2}} = a_0 + \delta a.
\ee

Finally, summing it all up, the expression for the dispersion relation in question is

\be 
s = s_0^0 + \delta s,
\ee

where $s_0^0$ is the solution of one of the two equations

\be  
\arcotanh s_0^0 = \frac{1}{s_0^0}\left(\frac{1}{2(F_S\pm F_A)} + 1\right),
\ee

and $\delta s$ is given by

\bea
\delta s = \mp(a_0+\delta a)^2 \frac{F_S L_2(s_0^0) + [L_1(s_0^0)^2 - 2L_0(s_0^0)L_2(s_0^0)](F_S^2-F_A^2)]}{F_A [\arcotanh s_0^0 - \frac{s_0^0}{2((s_0^0)^2 - 1)}] } \\
\approx \mp\frac{\omega^2\mu^2}{p_F^4}\left(1-\frac{2\pi^2T^2}{3v_F^2p_F^2}\right) \frac{F_S L_2(s_0^0) + [L_1(s_0^0)^2 - 2L_0(s_0^0)L_2(s_0^0)](F_S^2-F_A^2)]}{F_A [\arcotanh s_0^0 - \frac{s_0^0}{2((s_0^0)^2 - 1)}] }
\eea

So, we see that both at zero and non-zero temperature there are two modes of zero sound. In both cases the lowest order correction to the velocity of zero sound is proportional to the square of angular velocity. For the non-zero temperature case there is an extra correction, proportional to squares of both angular velocity and temperature. 

\section{Conclusion}

In this note we have discussed the mixing of the collective modes in the anomalous
kinetic equations in the cold Fermi liquid. We have reproduced the previous results
in the hydrodynamical limit and find a new modification of the anomalous zero sound
when the vorticity and temperature are taken into account.

The work of D.F. is supported by grant RFBR-15-02-02092.
The work of A.G. was performed at the Institute for
Information Transmission Problems with the financial support of the Russian Science Foundation
(Grant No.14-50-00150).

 \end{document}